\documentclass{iopart}
\usepackage{iopams}
\usepackage{epsfig}
\newcommand{\avg}[1]{\left\langle{#1}\right\rangle}

\newcommand{\ovl}[1]{\overline{#1}}
\renewcommand{\l}{\left}
\renewcommand{\r}{\right}
\begin{document}
\letter{Broken ergodicity and memory in the minority game} \author{J A
F Heimel\dag~and~A De Martino\ddag} \address{$\dag$ Dept. of
Mathematics, King's College London, Strand, London WC2R 2LS (UK)}
\address{$\ddag$ International School for Advanced Studies
(SISSA/ISAS) and INFM, via Beirut 2-4, 34014 Trieste (Italy)}
\eads{\mailto{heimel@mth.kcl.ac.uk}, \mailto{andemar@sissa.it}}
\begin{abstract}
We study the dynamics of the `batch' minority game with market-impact
correction using generating functional techniques to carry out the
quenched disorder average. We find that the assumption of weak
long-term memory, which one usually makes in order to calculate
ergodic stationary states, breaks down when the persistent
autocorrelation becomes larger than $c_c\simeq 0.772$. We show that
this condition, remarkably, coincides with the AT-line found in an
earlier static calculation. This result suggests a new scenario for
ergodicity breaking in disordered systems.
\end{abstract} 

The minority game\footnote[3]{See the web page
www.unifr.ch/econophysics/minority for an extensive and commented
overview of the existing literature.} (MG) models a market of
speculators interacting through a simple supply-and-demand mechanism
\cite{cz}. One of the key behavioural assumptions of the original
model is that agents act as so-called price-takers, meaning that at
every stage of the game each of them only perceives the aggregate
action of all agents, i.e. the total bid. Recently, in \cite{mcz}, a
generalization has been introduced in which agents are able to
estimate their own contribution to the total bid and use this
additional information to adjust their learning dynamics and optimize
their performance. The statics of this model has been tackled by
spin-glass techniques in \cite{mcz,dm} along the lines of
\cite{cmz}. The system was found to approximately minimize a
disordered hamiltonian $H$ whose minima could be calculated with the
replica method. It was shown that replica-symmetry breaking (RSB) can
occur, implying the existence of multiple stationary states.

In this Letter we adapt the dynamical method used in \cite{hc} to
analyze the `batch' version of this model. Assuming time-translation
invariance, finite integrated response and weak long-term memory
\cite{ck} we obtain exact results for the stationary state which are in
excellent agreement with computer experiments and with earlier static
approaches. Moreover, we derive a condition for the continuous onset
of memory, where the assumption of weak long-term memory is
found to fail while time-translation invariance still holds. This
appears to be different from the usual aging scenario in non-ergodic
disordered systems. Remarkably, the memory-onset condition
coincides with the AT line found in statics.

We begin by recalling the definition of the model. We consider $N$
agents labeled by roman indices. At each iteration round $n$ all
agents receive the same information pattern $\mu(n)$ drawn at random
with uniform probability from $\{1,\ldots,\alpha N\}$. Each agent has
at his disposal $S$ different strategies (labeled by $g=1,\ldots,S$)
to convert the acquired information into a trading
decision. Strategies are denoted by $\alpha N$-dimensional vectors:
$\bi{a}_{ig}=\{a_{ig}^\mu\}_{\mu=1}^{\alpha N}\in\{-1,1\}^{\alpha N}$,
where $a_{ig}^\mu$ is the trading action (e.g. $+1$ for `buy', $-1$
for `sell') prescribed to agent $i$ by his $g$-th strategy given
receipt of information $\mu$. By assumption, each component
$a_{ig}^\mu$ is selected randomly and independently from $\{-1,1\}$
with uniform probabilities before the start of the game, for all $i$,
$g$ and $\mu$. This introduces quenched disorder into the model.  Each
strategy of every agent is given an initial valuation $p_{ig}(0)$,
which is updated at the end of every round. At the start of round $n$,
given $\mu(n)$, every agent selects the strategy with the highest
valuation, $\widetilde{g}_i(n)=\arg\max p_{ig}(n)$, and subsequently
makes a bid according to the trading decision set by the selected
strategy: $b_i(n)=a_{i\widetilde{g}_i(n)}^{\mu(n)}$. The total bid at
round $n$ is defined as $A(n)=N^{-1/2}\sum_{i=1}^N b_i(n)$. Finally,
for all $i$ and $g$ all payoffs are updated according to a
reinforcement learning dynamics of the form
\begin{equation} 
p_{ig}(n+1)=p_{ig}(n)-a_{ig}^{\mu(n)}\l[A(n)-
\frac{\eta}{\sqrt{N}}\l(a_{i\widetilde{g}_i(n)}^{\mu(n)}-
a_{ig}^{\mu(n)}\r)\r]
\end{equation}
and agents move to the next round. The first term in square brackets
embodies the minority rule, in that the valuation of a strategy is
increased every time it predicts the correct minority action,
independently of it having been actually used. The term proportional
to $\eta$ adjusts the total bid for the possibility that agent $i$ is
not using strategy $\widetilde{g}_i(n)$. For $\eta=0$ one returns to
the original MG, while for $\eta=1$ the total bid is completely
adjusted.

We focus on the case $g=1,2$. Introducing the variables
$y_i(n)=[p_{i1}(n)-p_{i2}(n)]/2$, as well as the $\alpha
N$-dimensional vectors $\bomega_i=(\bi{a}_{i1}+\bi{a}_{i2})/2$,
$\bOmega=N^{-1/2}\sum_{i=1}^N\bomega_i$ and
$\bxi_i=(\bi{a}_{i1}-\bi{a}_{i2})/2$, and defining $s_i(n)={\rm
sgn}[y_i(n)]$ one has
\begin{equation}\label{ddd}
  y_i(n+1)
  =
  y_i(n)-\xi_i^{\mu(n)}[\Omega^{\mu(n)}+\frac{1}{\sqrt{N}}\!
  \sum_{j=1}^N\xi_j^{\mu(n)}s_j(n)-\frac{\eta}{\sqrt{N}}s_i(n)
  \xi_i^{\mu(n)}]
\end{equation}
Following \cite{hc} we study in this paper a `batch' version of the
model, which is obtained by averaging (\ref{ddd}) over information
patterns:
\begin{equation}\label{batch}
y_i(t+1)=y_i(t)-h_i-\sum_{j=1}^N J_{ij}s_j(t)+\eta\alpha s_i(t)+
\theta_i(t)
\end{equation}
where $t$ is a re-scaled time, $h_i=(2/\sqrt{N})~\bOmega\cdot\bxi_i$
and $J_{ij}=(2/N)~\bxi_i\cdot\bxi_j$. The external field $\theta_i(t)$
has been added for later use. In contrast to the more usual `on-line'
model (\ref{ddd}), where the $y_i$'s are updated after every iteration
step, in the `batch' case the updates are made on the basis of the
average effect of all possible choices of $\mu$. This modified
dynamics yields results for the stationary state which are
quantitatively very similar to those of the original model
\cite{ch}. The theoretical advantage of the `batch' formulation is
that it circumvents the difficulty of constructing a proper continuous
time limit. The numerical advantage is that one can simulate larger
systems for a longer time. Following \cite{hc}, one derives the
effective non-linear single-agent equation
\begin{equation}\label{dyna}
y(t+1)=y(t)-\alpha\sum_{t'\leq t}(\mathsf{I}+\mathsf{G})^{-1}_{tt'}
s(t')+ \alpha\eta s(t)+\sqrt{\alpha}z(t)+\theta(t)
\end{equation}
where $s(t)={\rm sgn}[y(t)]$ and $z(t)$ is a Gaussian noise with
zero mean and temporal correlations given by 
\begin{equation}\label{cortemp}
\avg{z(t)z(t')}\equiv H_{tt'}=\sum_{ss'}(\mathsf{I}+\mathsf{G})^{-1}_{ts}
(\mathsf{E}+\mathsf{C})_{ss'}(\mathsf{I}+\mathsf{G}^T)^{-1}_{s't'}
\end{equation}
The matrices $\mathsf{C}$ and $\mathsf{G}$ appearing here are the
noise-averaged single-agent correlation and response functions for the
process (\ref{dyna}), with elements
\begin{equation}\label{gc}
C_{tt'}=\avg{s(t)s(t')}\qquad{\rm and\qquad}
G_{tt'}=\avg{\frac{\partial s(t)}{\partial\theta(t')}}
\end{equation}
respectively, while $\mathsf{I}$ is the identity matrix and
$\mathsf{E}$ denotes the matrix with all entries equal to one. The
link between the Markovian multi-agent system (\ref{ddd}) and the
non-Markovian single-agent process (\ref{dyna}) is established by the
fact that, for $N\to\infty$, $C_{tt'}$ and $G_{tt'}$ become
identical to the disorder- and agent-averaged correlation and
response functions of (\ref{ddd}):
\begin{equation}
C_{tt'}=\frac{1}{N}\sum_{i=1}^N\l[s_i(t)s_i(t')\r]_{{\rm dis}}
\quad{\rm and}\quad
G_{tt'}=\frac{1}{N}\sum_{i=1}^N\l[\frac{\partial
s_i(t)}{\partial\theta_i(t')}\r]_{{\rm dis}}
\end{equation}

Eqs (\ref{dyna}-\ref{gc}) describe the dynamics of the system exactly
in the $N\to\infty$ limit. We now move to the stationary states of
(\ref{dyna}) upon making the following assumptions:

\noindent\hspace{-4pt}\begin{tabular}{ l@{~~} l }
Time-translation invariance (TTI) &
$\lim_{t\rightarrow\infty}C_{t+\tau,t}=C(\tau)~,~
\lim_{t\rightarrow\infty}G_{t+\tau,t}=G(\tau)$\\ Finite integrated
response (FIR) & $\lim_{t\rightarrow\infty}\sum_{t'\leq
t}G_{tt'}=\chi<\infty$ \\ Weak long-term memory (WLTM) &
$\lim_{t\to\infty}G(t,t')=0\quad\forall t'$ finite
\end{tabular}

\noindent For the re-scaled quantity $\widetilde{y}=\lim_{t\to\infty}y(t)/t$ one
finds
\begin{equation}
\widetilde{y}=-\frac{\alpha s}{1+\chi}+\alpha\eta s+\sqrt{\alpha}z+
\theta
\end{equation}
where $s=\lim_{\tau\to\infty}\tau^{-1}\sum_{t<\tau}{\rm sgn} [y(t)]$
and $z=\lim_{\tau\to\infty}\tau^{-1}\sum_{t<\tau}z(t)$, while $\theta$
is a static field. The variance of the zero-average Gaussian random
variable $z$ can be calculated from (\ref{cortemp}), yielding
\begin{equation}\label{var}
\avg{z^2}=\lim_{\tau,\tau'\to\infty}\sum_{t\leq\tau}\sum_{t'\leq\tau'}
H_{tt'}=\frac{1+c}{(1+\chi)^2}
\end{equation}
with the persistent correlation
$c\equiv\avg{s^2}=\lim_{\tau\to\infty}\tau^{-1}\sum_{t<\tau}C(t)$.
The effective agent is `frozen' if $\widetilde{y}\neq 0$, so that
$s={\rm sgn}(\widetilde{y})$ and he is always employing the same
strategy. Setting $\theta=0$, this is easily seen to be the case if
$|z|>\gamma$ with $\gamma=\sqrt{\alpha}\l[(1+\chi)^{-1}-\eta\r]$,
provided $\gamma\geq 0$. He is instead fickle when $\widetilde{y}=0$
or $|z|<\gamma$, and in this case $s=z/\gamma$. A self-consistent
equation for $c$ can now be derived by separating the contribution of
the frozen agents from that of the fickle ones. Upon defining
$\lambda=\gamma/\sqrt{\avg{z^2}}$ one finds
\begin{equation}\label{c}
c=\avg{\Theta(|z|-\gamma)}+\avg{\Theta(\gamma-|z|)
\frac{z^2}{\gamma^2}}= \phi+\frac{1}{\lambda^2}\l[\ovl{\phi}-\lambda
\sqrt{\frac{2}{\pi}}e^{-\frac{\lambda^2}{2}}\r]
\end{equation}
where $\Theta$ is the step function, $\ovl{\phi}={\rm
erf}(\lambda/\sqrt{2})$ is the fraction of fickle agents, and
$\phi=1-\ovl{\phi}$ is the fraction of frozen agents. For
$\chi=\avg{\frac{\partial s}{\partial
\theta}}=\alpha^{-1/2}\avg{\frac{\partial s}{\partial z}}$ one obtains
\begin{equation}\label{chi}
\chi=\frac{1}{\sqrt{\alpha}}\avg{\Theta(|z|-\gamma)2
\delta(\sqrt{\alpha}z)}+\frac{1}{\gamma\sqrt{\alpha}}
\avg{\Theta(\gamma-|z|)}=\frac{\ovl{\phi}}{\gamma\sqrt{\alpha}}
\end{equation}
Equations (\ref{var}-\ref{chi}) form a closed set from which one can
solve for $\phi$, $c$ and $\chi$ for any $\alpha$ and $\eta$. Results
for $c$ are shown in Figure 1.

For negative $\eta$, one observes an excellent agreement between
theory and experiment for all values of $\alpha$, implying that none
of our assumptions is ever violated.

When $\eta=0$, we recover the results of \cite{mcz,hc}, which match
the simulations perfectly for $\alpha$ larger than the critical value
$\alpha_c\simeq 0.3374$. At this point the integrated response $\chi$
diverges (FIR is violated) and a transition to a highly non-ergodic
regime takes place, where the stationary state depends on the initial
conditions $y(0)$. Starting with $y(0)\simeq 0$ leads to a high
volatility state, while starting with $|y(0)|\gg 1$ leads to
relatively low volatility.  The latter regime can be solved using the
assumption that $\chi$ remains very large for all
$\alpha<\alpha_c$. In fact, if $\chi\gg 1$ then
$\gamma\simeq\sqrt{\alpha}/\chi$ so that $\ovl{\phi}=\alpha$, which is
equivalent to ${\rm erf}(\lambda/\sqrt{2})=\alpha$. Solving this for
$\lambda$ and inserting the resulting value in (\ref{c}), we obtain
the top left branch of the $\eta=0$ curve in Fig. 1, which is again in
excellent agreement with numerical results.

For positive $\eta$, one sees that when $c>c_c\simeq 0.77$ our
theoretical predictions deviate from the experimental observations,
whereas the agreement is perfect for $c<c_c$. Finding no violation of
FIR, we have to conclude that either TTI or WLTM is violated. However,
we have found no evidence of aging. Therefore we expect the
deviations to be related to the breakdown of WLTM only. To find the
onset of memory, we split $G_{tt'}$ in its TTI part and its non-TTI
part:
\begin{equation}
\lim_{t\to\infty}G_{tt'}=\widetilde{G}(t-t')+\widehat{G}(t,t')
\end{equation}
During the initial stages of the game, small perturbations can cause
some agents, which would otherwise have remained fickle, to freeze and
vice versa, thus creating a persistent part $\mathsf{\widehat{G}}$ in
the response function. As the agents freeze, their state (and
consequently their contribution to $\mathsf{G}$) becomes independent
of $t$, so that we expect $\lim_{t\rightarrow\infty}\widehat{G}(t,t')
=\widehat{G}(t')$. After an initial equilibration period, for all
frozen agents the difference between the strategy valuations have
become very large, so they are virtually insensitive to
perturbations. Hence we must assume that
$\lim_{t'\to\infty}\widehat{G}(t')=0$. The fickle agents, however,
remain sensitive to small perturbations. The effects will wear out
over time (finite response) and are given by $\mathsf{\widetilde{G}}$.

Assuming $\mathsf{\widehat{G}}$ is small, we expand
$(\mathsf{I}+\mathsf{G})^{-1}$ in powers of $\mathsf{\widehat{G}}$ up
to first order:
\begin{equation}
  (\mathsf{I}+\mathsf{G})^{-1}
  =
  (\mathsf{I}+\mathsf{\widetilde{G}})^{-1}
  -\sum_{n=0}^\infty \sum_{m=0}^{n-1}
   (-\mathsf{\widetilde{G}})^m  \mathsf{\widehat{G}} 
   (-\mathsf{\widetilde{G}})^{n-m-1}
  +\Or(\mathsf{\widehat{G}}^2).
\end{equation}
Defining $\widetilde{\chi}=\sum_t\widetilde{G}(t)$ and
$\widehat{\chi}=\sum_t\widehat{G}(t)$, one then finds asymptotically
\begin{eqnarray}
  \widetilde{y}
  &=&
  -\alpha\l(\frac{1}{1+\widetilde{\chi}}-\eta\r)s
  +\sqrt{\alpha}z
\nonumber\\
  &&+\alpha \sum_{n=0}^{\infty}\sum_{m=0}^{n-1}
    (-\widetilde{\chi})^m 
    \sum_{t'} \widehat{G}(t')
    \sum_{t''} \left[(-\mathsf{\widetilde{G}})^{n-m-1}\right]\!(t',t'')s(t'')
\end{eqnarray}
Using the rectified linear function $f(x)=x$ for $|x|\leq 1$ and
${\rm sgn}(x)$ otherwise, we see that if
$1/(1+\widetilde{\chi})>\eta$ then
\begin{eqnarray}
\fl  s
  =
  f\l(\frac{1}{\widetilde{\gamma}}\l[
    z
    +\sqrt{\alpha} \sum_{n=0}^{\infty}\sum_{m=0}^{n-1}
    (-\widetilde{\chi})^m 
    \sum_{t'} \widehat{G}(t')
    \sum_{t''} \left[(-\mathsf{\widetilde{G}})^{n-m-1}\right]\!(t',t'')s(t'')
  \r]\r),
\\
  \widetilde{\gamma}=\sqrt{\alpha}\l(\frac{1}{1+\widetilde{\chi}}-\eta\r)
  \nonumber
\end{eqnarray}
As before, we have $\widetilde{\chi}=\alpha^{-1/2}\avg{\frac{\partial
s}{\partial z}}$, whereas
\begin{equation}
\fl  \widehat{G}(t)
  =\avg{\frac{\partial s}{\partial\theta(t)}}
  =
  \frac{\sqrt{\alpha}}{\widetilde{\gamma}}
  \sum_{n=0}^{\infty}\sum_{m=0}^{n-1}
    (-\widetilde{\chi})^m 
    \sum_{t'} \widehat{G}(t')
    \sum_{t''} \left[(-\mathsf{\widetilde{G}})^{n-m-1}\right]\!(t',t'')
  \widetilde{G}(t'',t)
\end{equation}
Up to first order in $\mathsf{\widehat{G}}$ one finds
$\widehat{\chi}=\widehat{\chi}\widetilde{\chi}\sqrt{\alpha}/[\widetilde{\gamma}
(1+\widetilde{\chi})^2]+\Or(\widehat{G}^2)$. Although
$\widehat{\chi}=0$ is always a solution of this equation, a
bifurcation occurs when
$\widetilde{\chi}\sqrt{\alpha}/[\widetilde{\gamma}
(1+\widetilde{\chi})^2]=1$, which is equivalent to $\ovl{\phi}=\alpha[1-\eta(1+\chi)]^2$, and can be written in terms of $\lambda$ as
\begin{equation} \label{mo}
\lambda^2[1+c(\lambda)]=\ovl{\phi(\lambda)}
\end{equation}
We call this line in the $(\alpha,\eta)$ plane the memory-onset (MO)
line, 
see Figure
2. It coincides remarkably with the AT-line (see Appendix), and
implies that the bifurcation occurs at $c_c\simeq 0.7722$ for
$\eta>0$. Above this value, WLTM can be broken, and indeed one sees
from Figure 1 that numerical results deviate from our theoretical
predictions for $c>c_c$. To give further evidence of memory, we have
analyzed the time evolution of two identical copies $a$ and $b$ of the
system, starting from slightly different initial conditions. We
plotted in Figure 3 the distance $d$ of the stationary states, given by
$(1/N)\sum_i(s_i^a -s_i^b)^2$, where $s_i^m$ is the long-time average
of ${\rm sgn}(y^m_i)$ ($m=a,b$), versus the persistent autocorrelation
of copy $a$, $c^a$. As $c^a$ approaches $c_c$, the two
copies end up in different stationary states, proving that they
remember initial conditions\footnote[4]{The slight bump that occurs
before $c_c$ is likely due to the fact that in our simulation the
perturbation can not be infinitesimal, but is at least $1/N$.}. At the
same time, if a perturbation is applied much later during the run the
copies end up in the same stationary state, indicating that indeed
$\widehat{G}(t')\to 0$ as $t'\to\infty$.

Summarizing, we have shown that in this model the usual connection
between broken ergodicity and broken TTI (aging), as seen for
instance in mean-field spin glasses \cite{bckm}, does not occur. In
contrast, we derive from the dynamics a condition for breakdown of
WLTM and continuous onset of memory within the TTI regime, which is
found to be equivalent to the AT-line found in the static
approach. This remarkable deviation from the well-known RSB/aging
picture is possibly due to the fact that the microscopic dynamics of
our model does not satisfy detailed balance.

\smallskip

\smallskip
 
We gratefully acknowledge support from and useful discussions with A C
C Coolen and M Marsili, and with S Franz. We also thank SISSA and
King's College London for reciprocal hospitality. This work originated
at the International Seminar on Statistical Mechanics of Information
Processing in Cooperative Systems (Dresden, March 2001).

\section*{Appendix}

While investigating the possible relation between our MO line and
replica-symmetry breaking, it became apparent that in the very final
step of the AT-line calculation in \cite{dm} a small error has
occurred. It was found that the replica-symmetric solution becomes
unstable when
\begin{equation}
\lim_{\beta\to\infty}\avg{\beta^2\l(\avg{s^2}-\avg{s}^2\r)^2}_z=
(1+\chi)^2
\end{equation}
where $\avg{f(s)}=Z_\beta(z)^{-1}\int_{-1}^1f(s)e^{-\beta V_z(s)}ds$
and $Z_\beta(z)=\int_{-1}^1 e^{-\beta V_z(s)}ds$, with
$V_z(s)=\frac{1}{2}\gamma s^2-zs$. The brackets $\avg{\cdots}_z$
denote a Gaussian average over $z$ having zero mean and variance
$\avg{z^2}_z=(1+q)/(1+\chi)^2$, $q$ being the overlap between two
different replicas (off-diagonal overlap matrix element). We have
absorbed a spurious factor $\sqrt{\alpha}$ in $\beta$. If we now
define $F(z)=-\lim_{\beta\to\infty}\beta^{-1}\log Z_\beta(z)$, the AT
line can be written as $\avg{F''(z)^2}_z=(1+\chi)^2$. By Laplace's
method we find $F(z)=V_z(s_0)$, $s_0$ being the minimum of $V_z$ in
$[-1,1]$. For $|z/\gamma|<1$, $s_0$ lies inside this interval and
$V_z(s_0)=z^2/(2\gamma)$, while for $|z/\gamma|>1$ $s_0$ is on the
border and $V_z(s_0)=\gamma/2-|z|$. This gives second derivatives that
are $-\gamma^{-1}$ and $0$, respectively. The AT-line is therefore
given by $\langle\gamma^{-2}\Theta(1-|z/\gamma|)\rangle_z+\langle
0~\Theta(|z/\gamma|-1)\rangle_z=(1+\chi)^2$. Recognizing the
non-vanishing term on the l.h.s. as the fraction $\ovl{\phi}$ of
fickle agents, we find
\begin{equation}\label{at}
\alpha[1-\eta(1+\chi)]^2=\ovl{\phi}
\end{equation}
similar to the result of \cite{dm} where in place of $\ovl{\phi}$ a
$1$ was reported. Written in terms of $\lambda$, the AT line is
identical to the MO line (\ref{mo}). We learned that 
the AT-line can also be derived
from the dynamical stability of Eqs. (\ref{ddd})\cite{m}.

\newpage

\begin{figure}
\begin{center}
~~~~~~~~\epsfig{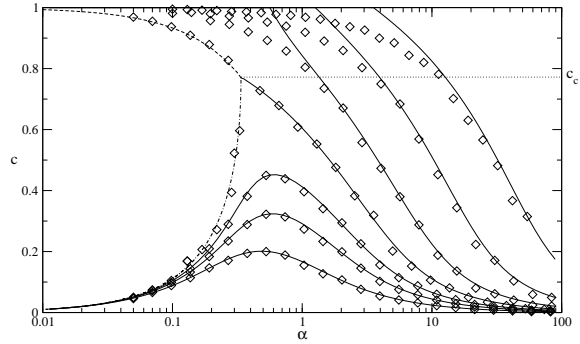}
\caption{The persistent correlation $c$ as a function of $\alpha$ for
different values of $\eta$. Lines represent theoretical predictions.
Solid lines: from bottom to top,
$\eta=-1,-0.5,-0.25,0,0.25,0.5,0.7$. Dashed line: $y(0)\gg
0,\eta=0$. Dot-dashed line: $\eta=0^-$. Diamonds correspond to
computer simulations with $\alpha N^2=10,000$, run for $500$ time
steps and averaged over $50$ disorder samples.}
\end{center}
\end{figure}

\begin{figure}
\begin{center}
\epsfig{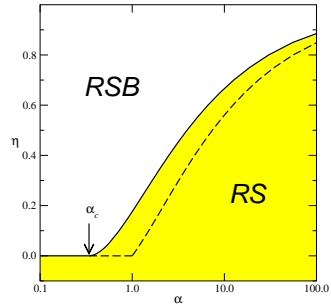}
\caption{The solid line represents the MO (=AT) line. The dashed line
corresponds to the AT line reported in \cite{dm}.}
\end{center}
\end{figure}

\begin{figure}[b]
\begin{center}
\epsfig{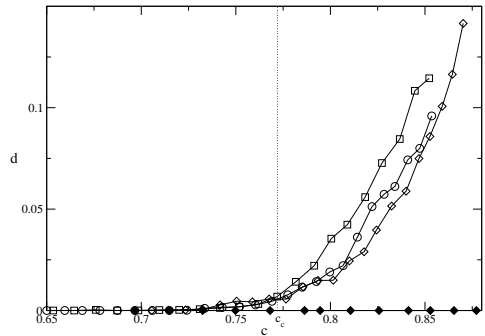}
\caption{Distance between the stationary states of two identical
copies of the system (see text) as a function of their persistent
autocorrelation. Simulations are for various levels of $\eta$ with
$N=450$, averaged over 100 samples. Open markers correspond to
a perturbation at $t=0$ ($\opencircle$, $\opensquare$, $\opendiamond$
for $\alpha=1,2,4$). Closed markers correspond to a perturbation at
$t=500$ ($\fulldiamond$ for $\alpha=2$). All simulations are run up to 500 steps after the perturbation occurs. Time averages are over the last 300 steps. }
\end{center}
\end{figure}

\end{document}